\documentstyle[11pt,newpasp,twoside,psfig]{article}
\index{X-rays}
\index{X-ray Instrumentation}

\def\edcomment#1{\iffalse\marginpar{\raggedright\sl#1\/}\else\relax\fi}
\reversemarginpar

\begin{document}
\title{High Resolution Optics \& Detector Systems for Hard X-rays}
 \author{D. P. Sharma, B. D. Ramsey, J. A. Gaskin, D. Engelhaupt, C. Speegle, 
 J. Apple \& M. H. Finger}
\affil{National Space Science and Technology Center, 320 Sparkman Drive, Huntsville, AL 35805}

\begin{abstract} We describe development of hard x-ray focusing
optics telescopes at the National Space Science and Technology Center (NSSTC).
\end{abstract}
High resolution grazing incidence optics has revolutionized
our understanding of
the universe at soft x-ray energies ($<$10 keV) as illustrated by the exciting
results from the Chandra Observatory (Weisskopf et al 2002). The hard x-ray
energy band still remains relatively unexplored at fine angular scales with 
high sensitivities due to the lack of such
technology.  Recent successful test flight (Ramsey et al. 2002) of the high 
spatial resolution and high sensitivity HERO
(High Energy Replicated Optics) payload has initiated a new era in hard
x-ray astronomy.

The Marshall Space Flight Center X-ray group at the NSSTC has an active program of
design, development and fabrication of replicated optics for hard X-ray
telescopes. The fabrication process involves generating super-polished mandrels,
mirror shell electroforming, and mirror testing. These mandrels are first
precision ground to within $\sim 1~\mu$m straightness along each conical segment and
then lapped and polished to $< 0.5~\mu$m straightness.  Each mandrel segment is then
super-polished to an average surface roughness of $\sim 4$\AA ~rms.  By mirror shell
replication, this combination of good figure and low surface roughness has
enabled us to achieve 15$\arcsec$ resolution in current mirror shells (Ramsey et al. 2003).

Currently, the focal plane detector for each mirror module is a high-pressure gas
scintillation proportional counter (Gubarev et al. 2003) with spatial resolution
of $\sim 400~\mu$m. However, the HERO optics with a focal spot diameter of around 430
$\mu$m for a 6 m focal length would require a spatial resolution of around 200 $\mu$m to
lead to a resolution of 15$\arcsec$. To match this resolution, we are developing fine
pixel CdZnTe detectors, each consisting of a CdZnTe crystal bonded to an ASIC readout chip.
Our current detectors have a 16x16 pixel array. We are evaluating two types
of ASICs: (a) an ASIC developed by Rutherford Appleton Laboratory, Oxford, England (RAL), 
which has preamplifiers for each pixel that output
to two other integrated circuits for shaping and peak detection; and (b) an ASIC developed at the
University of California, Riverside (UCR) which has an preamplifier, shaper, and peak detector
for each pixel on the same integrated circuit. The RAL detectors have pixel pad size of
250 $\mu$m with a pitch of 300 $\mu$m  whereas the UCR detectors have a pixel pad size of 
225 $\mu$m and a pitch of 300 $\mu$m. We are studying energy resolutions and spatial resolution
achievable, as well as charge loss and charge sharing between multiple pixels. Our current
evaluation program is aimed to lead us to the design and development of a 64x64
pixel array detector. 

\begin{figure}[t]
\centerline{\psfig{file=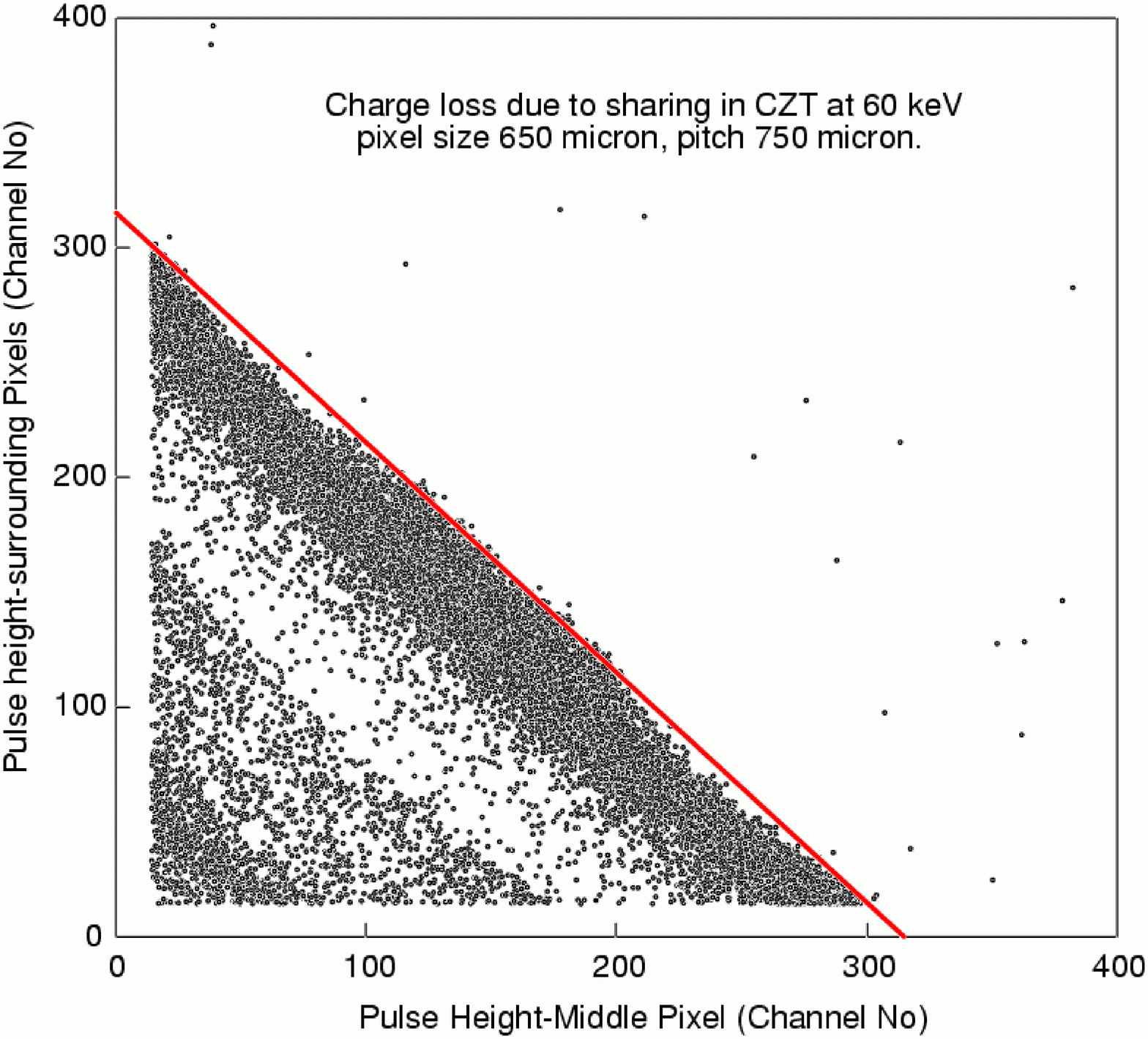,width=2.6in}\hspace{0.05in}
\psfig{file=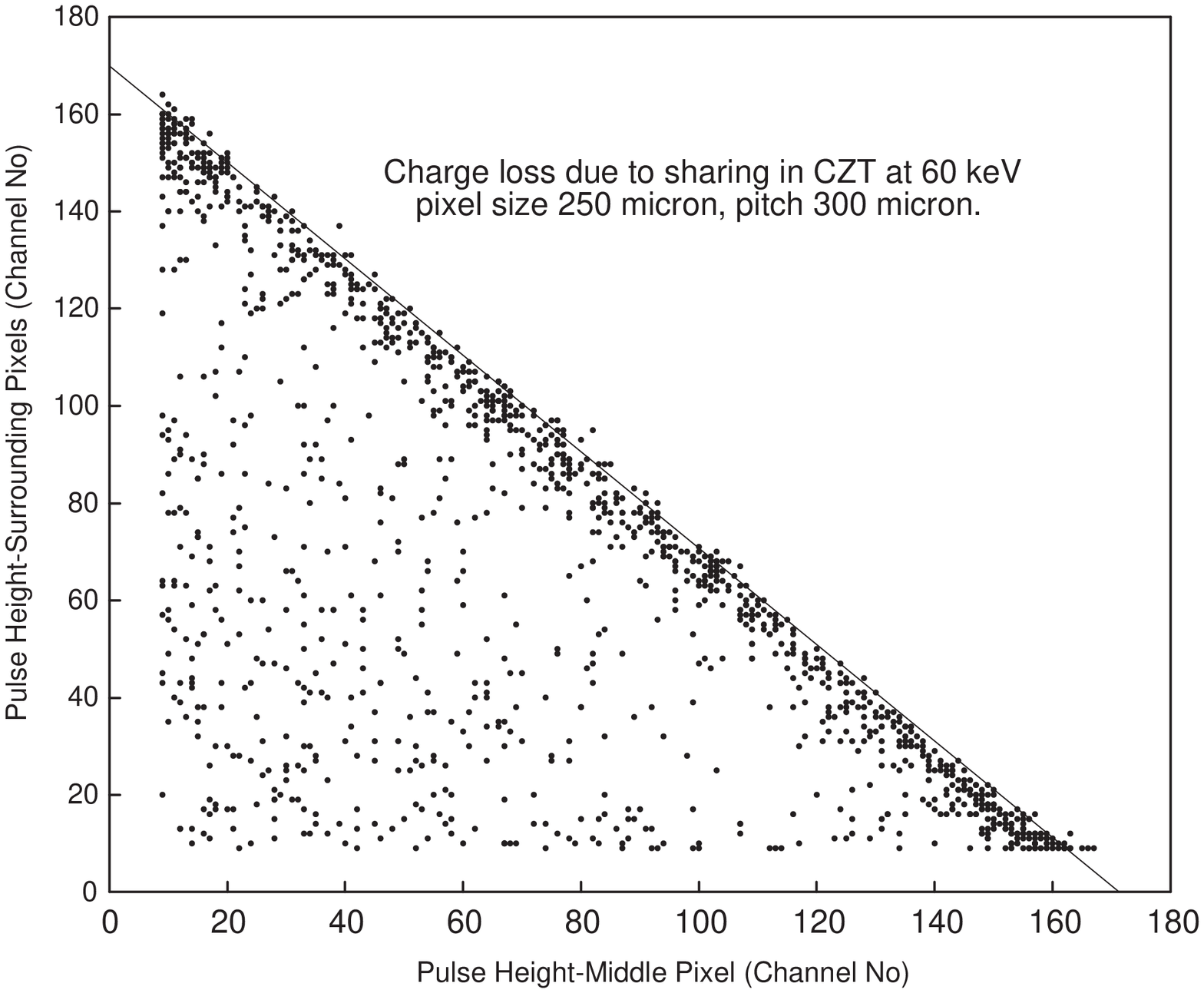,width=2.6in}}
\vspace*{-8pt}
\caption{The left panel shows events from a central pixel plotted against 
shared shared events from the four nearest surrounding pixels for a 
CdZnTe detector with 750 $\mu$m pixel spacing. The bow shape of the response is due to 
charge loss in shared events.
The right panel is for a detector with 300 $\mu$m pixel spacing, which shows
negligible charge loss in shared events (Sharma et al. 2002).}
\vspace*{-8pt}
\end{figure}

Charge sharing occurs when charge diffuses into adjacent pixels. The effect
depends on the pixel pitch and the amount the charge cloud spreads as it drifts
towards the anodes. We have measured this diffusion width to be around 42 $\mu$m,
and, for our 300-$\mu$m-pitch devices, this gives sharing of $\sim$ 50\% at 60 keV with a
2.5 keV threshold (the noise threshold plays a key role, as it sets the minimum
amount of shared charge that can be measured.) The down side of charge sharing
is that the output of multiple pixels must be added to recover the full charge
and this leads to an increase in the electronic noise and hence a worsening of
the energy resolution. A benefit of sharing, however, is that it permits
position interpolation which can provide better than pixel-pitch resolution
(Gaskin et al. 2003.)

Measured charge loss can be either due to a real loss of charge during
collection, or an apparent loss due to shared charge being below the noise threshold 
on a pixel. The former occurs mainly between pixels where the electric field is
low. It is particularly noticeable in devices with a large inter-pixel gap. The
latter occurs during sharing, and is simply dependent on the noise level on each
pixel.

\end{document}